\documentclass[12pt]{article}

\usepackage[a4paper, margin=2.5cm]{geometry}

\usepackage{cite}
\usepackage{longtable}
\usepackage{array}
\usepackage{booktabs}
\usepackage{ltxtable}
\usepackage{url}
\usepackage{hyperref}
\hypersetup{colorlinks=true, linkcolor=blue, citecolor=blue, urlcolor=blue}

\usepackage{amsmath,amssymb,amsfonts}
\newcommand{\tx}{\mathrm}

\usepackage{graphicx,color}
\usepackage{float}

\usepackage{siunitx}
\usepackage{multicol}

\usepackage{tikz}
\usetikzlibrary{arrows.meta,positioning}
\usepackage{pgfplots}
\pgfplotsset{compat=1.17}

\usepackage{subcaption}

\usepackage[font=small]{caption}

\usepackage{mathptmx}

\usepackage{textcomp}

\usepackage{algorithm}
\usepackage{algcompatible}

\usepackage{setspace}
\title{\textbf{Energy-Optimal Thermal Management of\\
Heat-Pump Battery Electric Vehicles}}

\author{
  Prashant Lokur$^{1,2}$,  Nikolce Murgovski$^{1}$\\[6pt]
  \normalsize $^{1}$Department of Electrical Engineering,
  Chalmers University of Technology, \\  \normalsize Göteborg 412~96, Sweden\\
  \normalsize $^{2}$Advanced Motion Systems \& Energy,
  Geely Technology Europe, Göteborg, Sweden\\[4pt]
  \normalsize Correspondence: \href{mailto:lokur@chalmers.se}{lokur@chalmers.se}
}

\date{%
  \small \textit{This paper has been submitted for publication in
  IEEE Open Journal of Vehicular Technology (OJVT).}\\[2pt]
  \small Preprint posted \today
}

\pgfplotsset{%
	compat=1.17,
	every non boxed x axis/.append style={x axis line style=-},
	every non boxed y axis/.append style={y axis line style=-},
	every non boxed z axis/.append style={z axis line style=-},
	every axis/.append style={%
		legend style={%
			fill=none,
			draw=none,
			align=left,
			legend cell align=left,
			font=\footnotesize,
			inner sep=0pt,
			outer sep=0pt,
			row sep=-2pt,
		},%
		xlabel near ticks,
		ylabel near ticks,
		label style={inner sep=0pt, outer sep=0pt},
		tick label style={inner sep=2pt,outer sep=0pt},
		x tick label style={inner xsep=0pt},
		y tick label style={inner ysep=0pt},
		axis lines = left,
		label style={font=\footnotesize},
		tick label style={font=\footnotesize},
		width=0.546\columnwidth,
		height=45mm
	}%
}

\begin{document}

\maketitle
\thispagestyle{empty}

% -----------------------------------------------------------------------
\begin{abstract}
This paper presents an energy-optimal hybrid control framework for thermal
management of heat-pump battery electric vehicles (BEVs). The controller
coordinates the compressor, coolant pumps, and cabin blower across the
coupled refrigerant, coolant, and air loops, while enforcing cabin comfort
and component temperature constraints. The framework combines a rule-based
supervisory layer, which handles discrete system configuration, with a
continuous nonlinear model predictive control (NMPC) optimizer that
minimizes thermal energy consumption over a finite prediction horizon. A
control-oriented model is developed to capture the dominant dynamics of the
cabin, refrigerant loop, reconfigurable coolant circuits, and key thermal
masses including the battery, motor, and inverter. The model is validated
against a high-fidelity reference, achieving a mean absolute temperature
prediction error below \SI{1.8}{\celsius} for key thermal states including
the battery, motor, and cabin air temperature, while reducing simulation
time by approximately \SI{85}{\percent}. The terminal cost is computed by
linearizing the system about a quasi-steady operating point and solving the
discrete-time algebraic Riccati equation, ensuring well-conditioned
optimization across varying operating conditions. The proposed framework is
evaluated against the built-in rule-based controller of the MathWorks
Simscape \emph{Electric Vehicle Thermal Management with Heat Pump} model
under cold-climate extended driving conditions, demonstrating consistent
reductions of \SI{20}{}--\SI{28}{\percent} in thermal energy consumption
across all tested scenarios. The complete implementation, developed using
the open-source CasADi framework, is made openly available at
\href{https://github.com/PrashantLokur/ThermalEnergyManagementWithHybridControlFramework}{GitHub}
to support reproducibility and further development.
\end{abstract}


\vspace{1em}
\hrule
\vspace{1em}

% -----------------------------------------------------------------------
%  Nomenclature
% -----------------------------------------------------------------------
\section*{Nomenclature}

\subsection*{Abbreviations}
\begin{tabular}{ll}
BEV   & Battery electric vehicle \\
COM   & Control-oriented model \\
DARE  & Discrete-time algebraic Riccati equation \\
EXV   & Expansion valve \\
HVAC  & Heating, ventilation, and air conditioning \\
MAE   & Mean absolute error \\
NLP   & Nonlinear program \\
NMPC  & Nonlinear model predictive control \\
NTU   & Number of transfer units \\
OCP   & Optimal control problem \\
RMSE  & Root mean square error \\
RK4   & Fourth-order Runge-Kutta \\
SOC   & State of charge \\
TEM   & Thermal energy management \\
WLTC  & Worldwide harmonized light vehicles test cycle \\
\end{tabular}

\subsection*{State Variables}
\begin{tabular}{ll}
$x$ & State vector \\
$T_{\tx{mot}}, T_{\tx{inv}}, T_{\tx{dcdc}}$ & Motor, inverter, and DC-DC temperatures (K) \\
$T_{\tx{b}}$ & Battery temperature (K) \\
$T_{\tx{int}}, T_{\tx{cair}}$ & Cabin interior mass and air temperatures (K) \\
$p_{\tx{in}}, p_{\tx{out}}$ & Compressor inlet and outlet pressures (Pa) \\
$\tx{SOC}$ & Battery state of charge (-) \\
\end{tabular}

\subsection*{Control Inputs}
\begin{tabular}{ll}
$u$ & Continuous control input vector \\
$\omega_{\tx{comp}}$ & Compressor speed (rpm) \\
$\dot{m}_{\tx{bl}}$ & Blower air mass flow rate (kg/s) \\
$\omega_{\tx{mot,p}}, \omega_{\tx{b,p}}$ & Motor-loop and battery-loop pump speeds (rpm) \\
$Q_{\tx{ht}}$ & Auxiliary heater power (W) \\
$\omega_{\tx{fan}}$ & Radiator fan speed (rpm) \\
\end{tabular}

\subsection*{Disturbances and Discrete Inputs}
\begin{tabular}{ll}
$d$ & Disturbance vector \\
$T_{\tx{amb}}$ & Ambient temperature (K) \\
$v_{\tx{veh}}$ & Vehicle speed (m/s) \\
$I_{\tx{b}}$ & Battery current (A) \\
$Q_{\tx{gen},i}$ & Heat generation rate of component $i$ (W) \\
$v$ & Discrete supervisory input vector \\
\end{tabular}

\subsection*{Key Variables}
\begin{tabular}{ll}
$i$ & Component index, $i \in \{\tx{b},\tx{mot},\tx{inv},\tx{dcdc}\}$ \\
$Q_{\tx{cool},i}$ & Heat removed from component $i$ (W) \\
$Q_{\tx{ab}}, Q_{\tx{rj}}$ & Net heat transfer at low- and high-pressure refrigerant sides (W) \\
$\dot{m}_{\tx{ref}}$ & Refrigerant mass flow rate (kg/s) \\
$\dot{m}_{\tx{b,clnt}},\dot{m}_{\tx{mot,clnt}}$ & Battery-loop and motor-loop coolant mass flow rates (kg/s) \\
$P_{\tx{TEM}}$ & Total TEM electrical power (W) \\
$UA_i$ & Overall heat transfer conductance of exchanger $i$ (W/K) \\
$\varepsilon_i$ & Effectiveness of exchanger $i$ (-) \\
$\Gamma_{\tx{ab}}, \Gamma_{\tx{rj}}$ & Thermodynamic capacitance coefficients (J/Pa) \\
$\bar{\phi}_{\tx{ab}}, \bar{\phi}_{\tx{rj}}$ & Mean void fractions at low- and high-pressure sides (-) \\
$h_1, h_2, h_{2s}$ & Compressor inlet, outlet, and isentropic outlet enthalpies (J/kg) \\
$R_{\tx{total}}$ & Total cabin envelope thermal resistance (K/W) \\
$T_{\tx{vent}}$ & HVAC supply air temperature (K) \\
\end{tabular}

\subsection*{Identified and Scheduled Parameters}
\begin{tabular}{ll}
$\boldsymbol{\gamma}$ & Vector of identified scaling parameters \\
$\boldsymbol{\gamma}(T_{\tx{amb}},x)$ & Scheduled parameter map \\
$\boldsymbol{\theta}$ & Thermophysical fluid property vector \\
\end{tabular}

\subsection*{NMPC Symbols}
\begin{tabular}{ll}
$N,\,\Delta t$ & Prediction horizon and sampling time \\
$\ell(\cdot),\,V_f$ & Stage and terminal cost \\
$x_{\tx{st}}, u_{\tx{st}}$ & Quasi-steady terminal state and input \\
$\hat{x}_t$ & State estimate at time $t$ \\
$s_k^{x,\ell},s_k^{x,u},s_k^y,s_k^{\Delta u}$ & Slack variables \\
$T_{\tx{cair}}^{\tx{ref}}$ & Cabin temperature reference (K) \\
$P$ & Terminal penalty matrix \\
$\gamma_d$ & DARE discount factor \\
\end{tabular}

\vspace{1em}
\hrule

% -----------------------------------------------------------------------
%  MAIN BODY — paste your sections here unchanged
% -----------------------------------------------------------------------

\section{Introduction}
\label{sec:intro}

Battery electric vehicles (BEVs) are a key pathway toward decarbonizing
road transport, supported by tightening emission regulations and the rapid
electrification of the passenger vehicles~\cite{StrictEnvironmentalRegulations}.
Despite significant advances in battery energy density and charging
infrastructure, limited driving range under real-world operating conditions
remains a significant barrier to customer acceptance and is commonly
referred to as range anxiety~\cite{RangeAnxietyStudies1,Pevec}.

The energy consumed by the thermal management system is a particularly
important contributor to this problem. Under harsh ambient conditions,
heating or cooling the cabin and keeping powertrain components within safe
temperature limits can draw substantial power from the traction battery,
leaving less energy available for propulsion. Field measurements indicate
that extreme weather can reduce BEV driving range by
\SIrange{30}{35}{\percent}~\cite{ThermalLoadImpact1}. Controlled testing at
Argonne National Laboratory documented even larger reductions for a Ford
Focus Electric Vehicle---\SI{53.7}{\percent} in hot weather and
\SI{59.3}{\percent} in cold weather on the UDDS
cycle~\cite{ThermalLoadImpact2}. In conventional vehicles, cabin heating
depends largely on engine waste heat at little additional energy cost; in
BEVs, both heating and cooling must be supplied by electrically driven
devices, making energy spent on thermal actuation a direct reduction in
driving range. Simply adding a larger battery can partially offset this,
but it adds weight and cost that make BEVs less competitive against
conventional vehicles. A more practical path is to minimize the energy
drawn by the thermal system itself---by controlling all thermal actuators
in a coordinated way that keeps passengers comfortable and all components
within safe operating limits. Achieving this in practice requires
understanding how modern BEV thermal architectures are structured.

Modern BEV thermal systems have moved well beyond simple resistive heaters
and single-loop cooling. Many production vehicles now combine
heat-pump-based heating, ventilation, air conditioning (HVAC) with
reconfigurable dual coolant circuits, which together enable waste-heat
recovery from the powertrain and improve heating efficiency significantly
at low ambient
temperatures~\cite{HeatPump}.
The refrigerant loop can run in two distinct modes: in cooling mode, heat
is rejected to the environment while the cabin air is cooled; in heat-pump
mode, heat is extracted from the ambient air, from powertrain waste heat,
or from both, and used to warm the cabin. At the same time, the dual
coolant circuits can be switched between series and parallel configurations
through directional valves, allowing waste heat from the motor, inverter,
and DC--DC converter to be routed to the battery pack, the cabin, or
rejected to the environment as needed.

All of these features are thermodynamically linked. Changing the
compressor speed, for example, affects cabin temperature, battery cooling,
and powertrain heat rejection all at once. A control strategy that tunes
one subsystem without considering the others will therefore miss efficiency
gains that are only accessible through coordinated
operation---and this becomes especially important in cold weather, when
thermal loads are at their heaviest.

Handling this kind of coupled, constrained problem is where model
predictive control (MPC) offers a structural advantage. By optimizing
actuator commands over a finite prediction horizon while explicitly
respecting system constraints, MPC can coordinate multiple thermal loops
in a manner that reactive or rule-based strategies are not readily able to
match~\cite{MPCforAutomotive1,MPCforAutomotive2}. Adaptive MPC
formulations have demonstrated real-time feasibility at automotive sampling
rates, including for energy-efficient torque distribution in heavy electric
vehicles~\cite{janardhanan2025energy}, motivating its application to the
more thermally complex setting addressed here. For nonlinear systems such
as BEV thermal management, nonlinear MPC (NMPC) is the appropriate
variant, though it requires a prediction model that captures the dominant
dynamics accurately enough to be useful, yet remains sufficiently simple to
evaluate repeatedly within the available sampling interval.
High-fidelity simulation models are generally too computationally expensive
and too nonsmooth to use directly inside an NMPC optimizer. There is also
the practical matter of discrete operating modes---the coolant circuits
switch between series and parallel configurations, and the refrigerant loop
switches between cooling and heat-pump operation---which makes a fully
mixed-integer formulation computationally intractable for real-time
deployment given the combinatorial complexity of solving such problems at
each sampling instant~\cite{bemporad1999control}. A further challenge
concerns the terminal cost, which in standard NMPC is typically designed
around a fixed operating point. For a thermal system that transitions
between heat-pump and cold-loop modes during a drive cycle, this assumption
breaks down. Near a mode transition, the linearized dynamics can be poorly
conditioned or marginally unstable, making a fixed terminal penalty
ill-suited and numerically unreliable exactly when thermal loads are
changing most rapidly. This paper addresses each of these challenges in
turn: a control-oriented nonlinear prediction model that remains tractable
for online optimization, a supervisory layer that resolves discrete mode
switching without integer variables, and a terminal cost that is updated
online to remain reliable across mode transitions.

Despite the clear need for integrated control, the literature has largely
addressed these challenges in isolation, treating the cabin, battery, and
powertrain as separate thermal problems with separate controllers. The
following review traces this fragmentation and identifies where the gaps
remain.

Cabin thermal comfort has received the most attention. A range of MPC
formulations have been proposed for cabin HVAC control---from approaches
that explicitly model passenger comfort and reduce online computation, to
strategies that handle ambient uncertainty and coordinate HVAC with
propulsion energy
management~\cite{schaut2019thermal,Dennis,XIE2021116084,9589323,Fei,jeffers2016climate}.
These consistently outperform rule-based controllers in energy use, but
they treat the cabin as an isolated system. The battery, motor, and power
electronics are either left out entirely or assumed to be regulated
separately, which means the controller cannot exploit interactions that are
fundamental to heat-pump architectures---such as using the compressor to
serve cabin and battery demands at the same time.

Battery and powertrain thermal management has been tackled in a similar
fashion. Optimization-based controllers have been developed to handle
uncertainty, enforce thermal limits, and account for long-term battery
degradation~\cite{9108617,7529090,BAUER2014808,WU2024122090}, and the
benefits in energy efficiency and component protection are well documented.
But in each case the refrigerant loop and cabin demands appear only as
external inputs, not as part of the optimization. In a heat-pump BEV,
where the compressor ties cabin conditioning and battery cooling together
through a shared circuit, this separation leaves real efficiency gains on
the table.

Some recent papers have started to bridge this gap by optimizing cabin and
battery thermal management jointly~\cite{hajidavalloo2023nmpc,CHEN202244}.
These are meaningful steps forward, but they stop short of including the
full set of powertrain thermal masses---the electric machine, inverter, and
DC--DC converter---whose coolant paths connect directly to the battery and
whose waste heat can meaningfully reduce cabin and battery heating demands
in cold conditions. Heat-pump mode transitions and the associated
refrigerant dynamics are also not modelled, which limits how well these
approaches transfer to the reconfigurable architectures found in production
vehicles. However, a practical issue that runs across much of this
literature is reproducibility: the high-fidelity plant models used for
evaluation are often proprietary or custom-built, making it difficult to
compare results across studies~\cite{lokur2024control}. Moreover,
in~\cite{lokur2024control}, the refrigerant pressure dynamics are
represented using an empirical approach that is valid only for a fixed
ambient temperature within a given scenario, and therefore requires
retuning when the ambient temperature changes. The present work addresses
all of these points---including the powertrain thermal masses, heat-pump
dynamics, and reproducible benchmarking---by evaluating the proposed
controller on the MathWorks Simscape \emph{Electric Vehicle Thermal
Management with Heat Pump}~\cite{MathWorksEVHeatPumpExample} and comparing
it against the built-in rule-based controller.

This paper presents an energy-optimal hybrid control architecture framework
for fully integrated BEV thermal management. The main contributions are as
follows.
\begin{itemize}
\item \textbf{Integrated control-oriented thermal energy management model:}
  A multi-domain model capturing the coupled dynamics of cabin air,
  refrigerant loop, reconfigurable coolant circuits, and key thermal
  masses---battery, motor, and inverter, and validated against a
  high-fidelity reference model.

\item \textbf{Constrained energy-optimal NMPC:} An NMPC formulation
  minimizing total electrical energy consumed by thermal
  actuators---compressor, pumps, and blower---subject to cabin comfort
  requirements, component temperature and pressure limits, and actuator
  bounds and rate constraints, with soft constraints to ensure feasibility
  under transient conditions.

\item \textbf{Adaptive terminal cost with robust Riccati computation:} At
  each sampling instant or when operating conditions change significantly,
  a quasi-steady terminal target $(x_\tx{st}, u_\tx{st})$ is obtained by
  solving a small nonlinear program (NLP) and the terminal penalty matrix
  $P$ is computed by solving the discrete-time algebraic Riccati equation.

\item \textbf{Hybrid discrete--continuous control structure:} A rule-based
  supervisory layer determines discrete mode selections at each sampling
  instant and holds them fixed over the prediction horizon, reducing the
  online NMPC to a continuous nonlinear program solved using warm-started
  interior-point optimization.

\item \textbf{Simulation-based evaluation:} Assessment of the proposed
  NMPC against a high-fidelity reference model and a rule-based baseline
  under cold-climate WLTC driving, covering thermal energy consumption,
  cabin comfort tracking, constraint satisfaction, and solver computational
  performance.
\end{itemize}

The remainder of this paper is organized as follows.
Section~\ref{sec:SystemArchitecture} describes the BEV thermal system
architecture. Section~\ref{sec:modeling} presents the control-oriented
dynamic model. Section~\ref{sec:ParameterTuning} details the model
parameter identification procedure. Section~\ref{sec:ModelValidation}
reports the model validation results against the high-fidelity reference.
Section~\ref{sec:nmpc} formulates the constrained energy-optimal NMPC
problem, including the hybrid control architecture and terminal cost
design. Section~\ref{sec:Results} presents the simulation results and
discussion. Section~\ref{sec:conclusion} concludes the paper.

% -----------------------------------------------------------------------
\section{System Architecture}
\label{sec:SystemArchitecture}

\begin{figure*}[t]
    \centering
    \includegraphics[width=\textwidth,height=0.34\textheight,keepaspectratio]
      {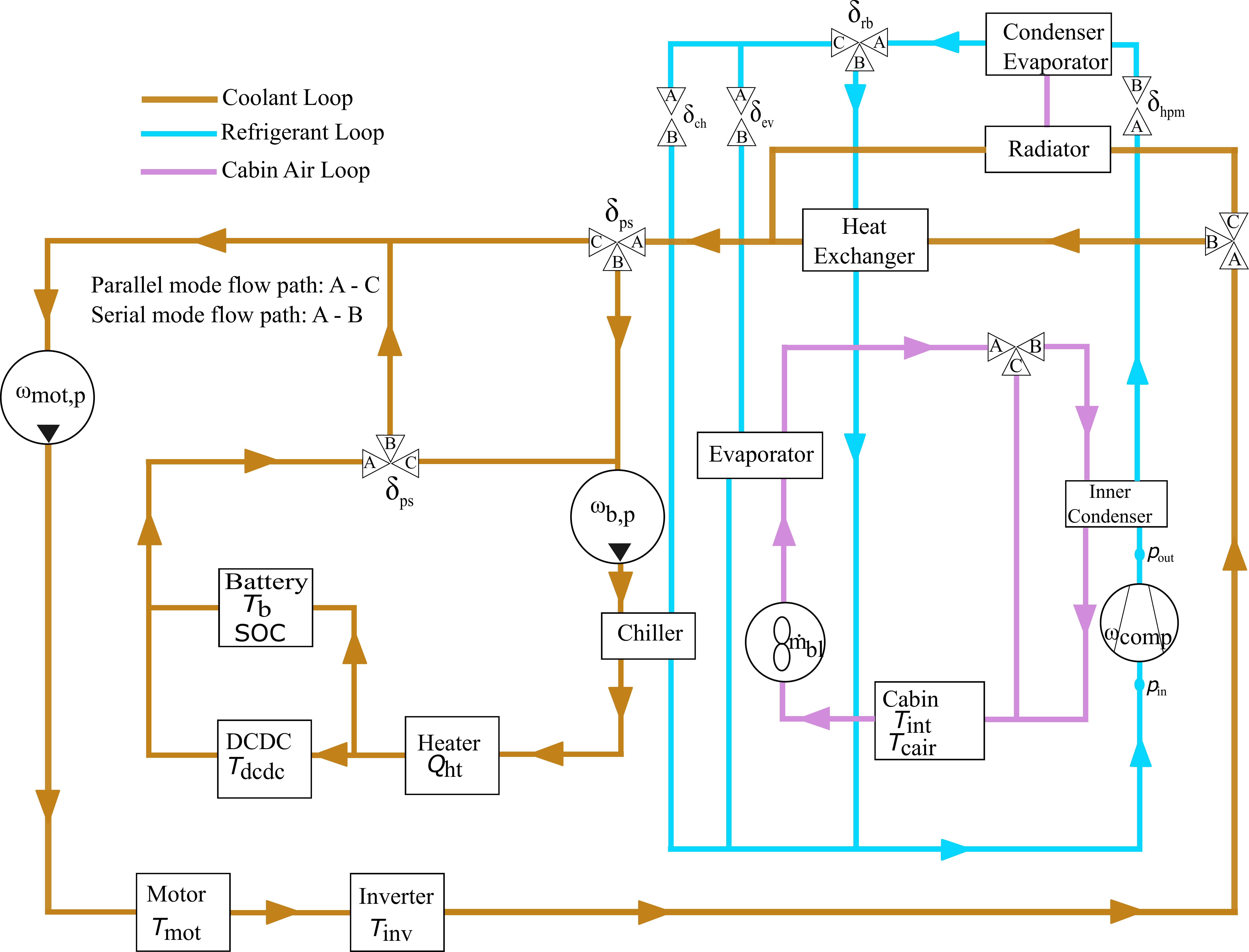}
    \caption{Schematic of the BEV thermal energy management (TEM)
      architecture considered in this work, based on the MathWorks
      Simscape \emph{Electric Vehicle Thermal Management with Heat Pump}
      model. Three coupled fluid loops are shown: the dual coolant loop
      (brown), the refrigerant loop (cyan), and the cabin air loop (pink).
      The coolant loop can be configured in series (path A--B) or parallel
      (path A--C) via directional valves. Continuous control inputs include
      the compressor speed $\omega_{\tx{comp}}$, pump speeds
      $\omega_{\tx{mot,p}}$ and $\omega_{\tx{bt,p}}$, blower flow rate
      $\omega_{\tx{bl}}$, and auxiliary heater power $Q_{\tx{ht}}$. The
      binary variables $\delta_{\tx{hpm}}$, $\delta_\tx{ps}$,
      $\delta_{\tx{rb}}$, $\delta_{\tx{ev}}$, and $\delta_{\tx{ch}}$
      denote the discrete mode-selection flags. Monitored states include
      component temperatures $T_{\tx{mot}}$, $T_{\tx{inv}}$,
      $T_{\tx{dcdc}}$, $T_{\tx{b}}$, cabin temperatures $T_{\tx{int}}$
      and $T_{\tx{cair}}$, battery SOC, and refrigerant pressures
      $p_{\tx{in}}$ and $p_{\tx{out}}$.}
    \label{fig:TEM_schematic}
\end{figure*}

The TEM architecture considered in this work is shown in
Fig.~\ref{fig:TEM_schematic}. It consists of three coupled fluid
loops---a dual coolant circuit, a refrigerant loop, and a cabin air
loop---whose configurations are switched depending on ambient temperature
and thermal demand. The following subsections describe each loop and its
operating modes.

\subsection{Coolant Circuits}
The two liquid coolant loops can be configured in either series or parallel
mode through directional valves. In cold conditions, the loops operate in
series, routing waste heat from the electric machine, inverter, and DC--DC
converter to the battery pack to support warm-up. Depending on the
operating mode, this heat can also be exchanged with the refrigerant
through a heat exchanger. When recovered waste heat is insufficient, an
auxiliary heater provides supplementary thermal power to the coolant.
Under mild ambient conditions, the radiator alone is sufficient to reject
heat from both the battery and the power electronics. Above approximately
\SI{35}{\celsius}, the system switches to parallel operation: one loop
rejects powertrain heat through the radiator independently while the other
cools the battery pack through a chiller coupled to the refrigerant loop.

\subsection{Refrigerant Loop}
The refrigerant loop operates in two distinct modes. In cold loop mode,
the cabin air bypasses the inner condenser and the heat pump expansion
valve remains fully open. The condenser/evaporator rejects heat to the
environment; the chiller expansion valve cools the coolant loop; and the
evaporator expansion valve cools the cabin air directly. In heat pump
mode, which is activated under cold ambient conditions, the inner condenser
is used to heat the cabin air. The condenser/evaporator transitions to
acting as an evaporator, absorbing heat from the ambient environment. The
heat pump expansion valve meters the refrigerant flow, and a bypass around
the evaporator and chiller directs refrigerant to a heat exchanger that
extracts additional waste heat from the powertrain coolant circuit,
improving overall heating efficiency.

The strong couplings between these loops---shared actuators, reconfigurable
flow paths, and mode-dependent heat transfer routes---motivate a
coordinated control approach that can optimize all actuators jointly, as
described in the following sections.

% -----------------------------------------------------------------------
\section{Dynamic Modeling of the BEV Thermal Management System}
\label{sec:modeling}

A control-oriented model (COM) of the BEV thermal management system is
developed to capture the dynamics of the cabin, refrigerant loop, coolant
circuits, and main powertrain components in a form suitable for
gradient-based optimization. Time dependence is omitted for readability,
i.e., $x(t)$ is written as $x$. Constant parameters are denoted with an
overbar, e.g., $\bar{C}_{\tx{nom}}$.

The COM dynamics are expressed as
$\dot{x} = f(x,\, u,\, d,\, v,\, \boldsymbol{\gamma},\, \boldsymbol{\theta})$
where $x$ is the state vector, $u$ is the continuous control input, $d$ is
the disturbance vector, and $v$ represents the discrete inputs determined
by the supervisory layer and held fixed over the prediction horizon
(Section~\ref{sec:nmpc}). The vector $\boldsymbol{\gamma}$ collects the
identified parameters (Section~\ref{sec:ParameterTuning}), and
$\boldsymbol{\theta}$ collects thermophysical fluid properties evaluated at
the current operating point via CoolProp~\cite{Bell2014} and passed as
fixed parameters over the prediction horizon.

The state vector is defined as
\begin{equation}
x = \begin{bmatrix}
T_\tx{mot} & T_\tx{inv} & T_\tx{dcdc} & \tx{SOC} & T_\tx{b} &
p_{\tx{in}} & p_{\tx{out}} & T_\tx{int} & T_\tx{cair}
\end{bmatrix}^\top
\label{eq:state}
\end{equation}

The continuous control input vector optimized by the NMPC is
\begin{equation}
u = \begin{bmatrix}
\omega_\tx{comp} & \dot{m}_\tx{bl} & \omega_\tx{mot,p} &
\omega_\tx{b,p} & Q_\tx{ht} & \omega_{\tx{fan}}
\end{bmatrix}^\top
\label{eq:input}
\end{equation}

The disturbance vector is defined as
\begin{equation}
d = \begin{bmatrix}
T_{\tx{amb}} & v_{\tx{veh}} & I_{\tx{b}} & Q_{\tx{gen,mot}} &
Q_{\tx{gen,dcdc}} & Q_{\tx{gen,inv}}
\end{bmatrix}^\top
\end{equation}

\subsection{Traction Power and Component Losses}
The component heat generation rates $Q_{\tx{gen},i} \in d$ are estimated
from lookup tables parameterized by vehicle speed and electrical operating
conditions, identified from the reference model.

\subsection{Battery Electrical--Thermal Model}
The battery pack is represented using a simple equivalent-circuit model
consisting of an open-circuit voltage source in series with an internal
resistance $R_{\tx{b}}$. The heat generation rate is approximated as
\begin{equation}
Q_{\tx{gen,b}} = I_{\tx{b}}^2 R_{\tx{b}}(T_{\tx{b}}, \tx{SOC})
\label{eq:bat_heat}
\end{equation}
where battery internal resistance $R_{\tx{b}}$ is modeled as a
second-order polynomial in battery temperature and state of charge,
\begin{equation}
R_{\tx{b}}(T_{\tx{b}}, \tx{SOC}) =
\sum_{i=0}^{2} \sum_{j=0}^{2} \bar{\psi}_{ij}\, T_{\tx{b}}^{i}\, \tx{SOC}^{j}
\label{eq:rb_poly}
\end{equation}
The SOC dynamics follow Coulomb counting,
\begin{equation}
\frac{\tx{d}\,\tx{SOC}}{\tx{d}t} = -\frac{I_{\tx{b}}}{\bar{C}_{\tx{nom}}}
\end{equation}

\subsection{Coolant Flow and Pump Models}
The coolant mass flow rate delivered by the battery coolant pump is modeled
using a fixed-displacement volumetric formulation,
\begin{align}
\dot{m}_{\tx{b,clnt}} = {\rho}_{\tx{clnt}} \cdot \bar{\alpha}_{\tx{p}} \cdot
\bar{\eta}_{\tx{vol}} \cdot \bar{V}_{\tx{disp,b}} \cdot
\frac{\omega_{\tx{b,p}}}{60}
\label{eq:clntmf}
\end{align}
The pressure drop across each circuit is approximated as quadratic in mass
flow rate,
\begin{align}
\Delta p_{\tx{b}} = \bar{k}_\tx{b} \cdot \dot{m}_{\tx{b,clnt}}^2
\label{eq:pdrop}
\end{align}
The corresponding pump electrical power is
\begin{align}
P_{\tx{b,pump}} = \frac{\dot{m}_{\tx{b,clnt}} \cdot \Delta p_{b}}
{{\rho}_{\tx{clnt}} \cdot \bar{\eta}_{\tx{p}}}
\label{eq:pumppwr}
\end{align}

\subsection{Heat Exchanger Framework}
\label{sec:hx}
All heat exchangers in the COM are modeled using the NTU-effectiveness
method. For each fluid stream
$s\in \{\tx{air, coolant, refrigerant}\}$, the Reynolds and Prandtl
numbers are
\begin{equation}
\mathrm{Re}_s = \frac{\rho_s\, v_s\, \bar{D}_s}{\mu_s}, \qquad
\mathrm{Pr}_s = \frac{c_{p,s}\, \mu_s}{k_s}
\label{eq:re_pr}
\end{equation}
The Nusselt number is obtained from the Dittus-Boelter
correlation~\cite{cengel2003heat},
$\mathrm{Nu}_s = 0.023\, \mathrm{Re}_s^{0.8}\, \mathrm{Pr}_s^{1/3}$,
giving the convective heat transfer coefficient
\begin{equation}
h_s = \frac{\mathrm{Nu}_s\, k_s}{\bar{D}_s}
\label{eq:htc}
\end{equation}
The overall heat transfer conductance is
\begin{align}
\frac{1}{UA} &= \left(\frac{1}{h_{s,\tx{hot}}\, \bar{A}} +
\frac{1}{h_{s,\tx{cold}}\, \bar{A}}\right)
\label{eq:ua}
\end{align}
The number of transfer units and exchanger effectiveness are
\begin{align}
\mathrm{NTU} &= \frac{UA}{C_{\min}},\qquad
\varepsilon = 1 - \exp(-\mathrm{NTU})
\label{eq:effectiveness}\\
Q &= \varepsilon C_{\min}(T_{s,\tx{hot,in}} - T_{s,\tx{cold,in}})
\label{eq:hx_q}
\end{align}
The outlet temperatures of the hot and cold streams are computed from
energy balances as
\begin{align}
T_{s,\tx{hot,out}} &= T_{s,\tx{hot,in}} -
\frac{Q}{\dot{m}_{s,\tx{hot}}\, c_{p,\tx{hot}}} \label{eq:htxTempOut}\\
T_{s,\tx{cold,out}} &= T_{s,\tx{cold,in}} +
\frac{Q}{\dot{m}_{s,\tx{cold}}\, c_{p,\tx{cold}}}
\end{align}

\subsection{Auxiliary Heater and Fan}
The coolant outlet temperature after the auxiliary heater is
\begin{equation}
T_{\tx{clnt,out,ht}} = T_{\tx{clnt,in,ht}} +
\frac{\bar{\eta}_{\tx{ht}} \cdot Q_{\tx{ht}}}
{\bar{\alpha}_{\tx{ht}} \cdot \dot{m}_{\tx{b,clnt}} \cdot c_{p,\tx{b,clnt}}}
\label{eq:heater}
\end{equation}
The electrical power consumed by the radiator fan is
\begin{equation}
P_{\tx{fan}} = \frac{\bar{P}_{\tx{fan,nom}}}{\bar{\eta}_{\tx{fan}}}
\left(\frac{\omega_{\tx{fan}}}{\bar{\omega}_{\tx{fan,ref}}}\right)^3
\label{eq:fan_power}
\end{equation}
The air mass flow through the condenser/evaporator combines ram air and
forced convection:
$\dot{m}_{\tx{ce,air}} = \bar{\alpha}_{\tx{ram}} \cdot v_{\tx{veh}}
+ \bar{\alpha}_{\tx{fan}} \cdot \omega_{\tx{fan}}$.

\subsection{Component Thermal Models}
The temperature dynamics of component $i \in \{\tx{b}, \tx{inv},
\tx{dcdc}, \tx{mot}\}$ are governed by
\begin{align}
\bar{m}_i \bar{c}_{p,i} \frac{\tx{d}T_i}{\tx{d}t} =
\gamma_i \left(Q_{\tx{gen},i} - Q_{\tx{cool},i}\right)
\label{eq:lumped_mass}
\end{align}
The heat removed from each component is
\begin{align}
Q_{\tx{cool},i} =& \varepsilon_i\, \dot{m}_{c,\tx{clnt}}\, c_{p,\tx{clnt}}
\,(T_i - T_{\tx{clnt,in},i}) \nonumber\\
&+ \bar{\kappa}_{\tx{cond},i}\,
\frac{\bar{A}_{\tx{hx}}}{\bar{D}_{\tx{ch}}}\,(T_i - T_{\tx{clnt,in},i})
\label{eq:q_cool}
\end{align}

\subsection{Refrigerant Loop}
The volumetric efficiency of the compressor is
\begin{align}
\eta_{\tx{v}} = \bar{\alpha}_{\tx{v}} \frac{p_{\tx{out}}}{p_{\tx{in}}} +
\bar{\beta}_{\tx{v}}
\label{eq:vol_eff}
\end{align}
The refrigerant mass flow rate is
\begin{align}
\dot{m}_{\tx{ref}} =
\frac{\eta_{\tx{v}} \, \omega_{\tx{comp}} \, \bar{V}_{\tx{disp,comp}} \,
\bar{\alpha}_{\tx{mf}}}{60\,{v}_{\tx{in}}}
\label{eq:comp_mass_flow}
\end{align}
The compressor outlet enthalpy and electrical power are
\begin{align}
h_2 &= h_1 + \frac{h_{2s} - h_1}{\bar{\eta}_{\tx{isen}}}
\label{eq:comp_enthalpy}\\
P_{\tx{comp}} &= \frac{\dot{m}_{\tx{ref}} (h_2 - h_1)}
{\bar{\eta}_{\tx{mech}} \cdot \bar{\eta}_{\tx{elec}}}
\label{eq:comp_power}
\end{align}
The pressure state equations are derived from a mean void fraction
model~\cite{zhang2015energy},
\begin{subequations}\label{eq:pressure_dynamics}
\begin{align}
\Gamma_{\tx{ab}} \frac{dp_{\tx{in}}}{dt} &=
\gamma_{6}\left(Q_{\tx{ab}} + \dot{m}_{\tx{ref}}(h_4 - h_1)\right)
\label{eq:evap_pressure_dynamics}\\
\Gamma_{\tx{rj}} \frac{dp_{\tx{out}}}{dt} &=
\gamma_{7}\left(-Q_{\tx{rj}} + \gamma_8 \dot{m}_{\tx{ref}}(h_2 - h_3)\right)
\label{eq:cond_pressure_dynamics}
\end{align}
\end{subequations}
The net heat transfer terms are
\begin{align}
Q_{\tx{ab}} &= \delta_{\tx{hpm}} Q_{\tx{ce}} + \delta_{\tx{rb}} Q_{\tx{hx}}
+ \delta_{\tx{ev}} Q_{\tx{ev}} + \delta_{\tx{ch}} Q_{\tx{ch}} \\
Q_{\tx{rj}} &= \delta_{\tx{w}} Q_{\tx{ic}} +
(1 - \delta_{\tx{hpm}}) Q_{\tx{ce}}
\end{align}
The thermodynamic capacitance coefficients $\Gamma_{\tx{ab}}$ and
$\Gamma_{\tx{rj}}$ are given in~\eqref{eq:Gab}--\eqref{eq:Grj}.
\begin{align}
\Gamma_{\tx{ab}} &= \bar{V}_{\tx{ab}} \Bigg[
(1-\bar{\phi}_{\tx{ab}}) \frac{\partial(\rho_{\tx{l}} h_{\tx{l}})}{\partial p_{\tx{in}}}
+ \bar{\phi}_{\tx{ab}} \frac{\partial(\rho_{\tx{g}} h_{\tx{g}})}{\partial p_{\tx{in}}}
- 1 + \frac{\bar{M}_{\tx{wab}} \bar{C}_{\tx{ab}}}{\bar{V}_{\tx{ab}}}
\frac{\partial T_{\tx{lp,sat}}}{\partial p_{\tx{in}}} \Bigg]
\label{eq:Gab}\\
\Gamma_{\tx{rj}} &= \bar{V}_{\tx{rj}} \Bigg[
(1-\bar{\phi}_{\tx{rj}}) \frac{\partial(\rho_{\tx{l}} h_{\tx{l}})}{\partial p_{\tx{out}}}
+ \bar{\phi}_{\tx{rj}} \frac{\partial(\rho_{\tx{g}} h_{\tx{g}})}{\partial p_{\tx{out}}}
- 1 + \frac{\bar{M}_{\tx{wrj}} \bar{C}_{\tx{rj}}}{\bar{V}_{\tx{rj}}}
\frac{\partial T_{\tx{hp,sat}}}{\partial p_{\tx{out}}} \Bigg]
\label{eq:Grj}
\end{align}
The compressor outlet temperature is approximated as
\begin{align}
T_{\tx{ref,out,comp}} = T_{\tx{sat,hp}} +
\bar{\alpha}_\tx{sh} \frac{h_{2s}-h_1}{c_{p,\tx{ref,vh}}}
\label{eq:comp_out_temp}
\end{align}

\subsection{Cabin Thermal Modeling}
A two-node lumped parameter model represents the cabin thermal dynamics.
The thermal resistance of each envelope element
$k \in \{\tx{glass}, \tx{doors}, \tx{roof}\}$ is
\begin{align}
R_k = \bar{\beta}_k \left( \frac{1}{\bar{U}_k \bar{A}_k} +
\frac{\bar{\delta}_k}{\bar{\lambda}_k \bar{A}_k} \right)
\label{eq:Rk}
\end{align}
The cabin interior mass temperature dynamics are
\begin{align}
\bar{M}_{\tx{int}} \bar{c}_{p,\tx{int}} \frac{dT_{\tx{int}}}{dt} =
\gamma_9 \left(\frac{T_{\tx{amb}} - T_{\tx{int}}}{R_{\tx{total}}} +
\bar{\alpha}_{\tx{int}} \frac{T_{\tx{cair}} - T_{\tx{int}}}{R_{\tx{total}}}\right)
\label{eq:cabin_int_ode}
\end{align}
The cabin air temperature dynamics are
\begin{equation}
\begin{split}
C_{\tx{air}} \frac{\tx{d}T_{\tx{cair}}}{\tx{d}t} = \gamma_{10}
\Bigg(&\dot{m}_{\tx{bl}} \bar{c}_{p,\tx{air}}
(T_{\tx{vent}} - T_{\tx{cair}}) \\
&+ \bar{Q}_{\tx{human}}
+ \frac{T_{\tx{int}} - T_{\tx{cair}}}
{\bar{\alpha}_{R,\tx{int}}\, R_{\tx{total}}}\Bigg)
\end{split}
\label{eq:cabin_air_ode}
\end{equation}
where the HVAC supply air temperature is
\begin{align}
T_{\tx{air,out},ic} &= T_{\tx{air,in},ic} +
\frac{Q_\tx{ic}}{\dot{m}_{\tx{bl}} c_{p,\tx{air}}}
\label{eq:innerCondAirOutlet}\\
T_{\tx{air,in},ic} &= (1-\bar{r}_{\tx{rec}})\;T_\tx{amb} +
{\bar{r}_\tx{rec}}T_{\tx{cair}}
\end{align}

\subsection{Total Actuator Power}
\begin{equation}
P_\tx{TEM} = P_\tx{comp} + P_\tx{bl} +
P_\tx{mp,pump} + P_\tx{bp,pump} +
Q_\tx{ht} + P_\tx{fan}
\label{eq:p_tem}
\end{equation}

% -----------------------------------------------------------------------
\section{Model Parameter Identification}
\label{sec:ParameterTuning}

The control-oriented model contains several simplifying assumptions
introduced to make the model suitable for real-time optimization. The
unknown parameters are therefore identified offline so that the resulting
model remains representative of the reference dynamics over the operating
conditions considered. The COM with parameter scheduling is written as
\begin{equation}
\dot{x} = f\bigl(x,\, u,\, d,\, v,\,
\boldsymbol{\gamma}(T_{\tx{amb}},x,I_\tx{b}),\, \boldsymbol{\theta}\bigr)
\label{eq:com_dynamics}
\end{equation}
The high-fidelity reference model is
\begin{equation}
\dot{x}_{\tx{M}} = f_{\tx{M}}(x_{\tx{M}},\, u_{\tx{M}},\, d)
\label{eq:ref_dynamics}
\end{equation}
Reference simulations are performed at
$T_{\tx{amb}} \in \{\SI{-10}{\celsius},\,\SI{-5}{\celsius}\}$.
For each window~$j$, the local parameter vector $\boldsymbol{\gamma}^{(j)}$
is identified by minimizing the weighted prediction error:
\begin{subequations}
\label{eq:param_id}
\begin{align}
\min_{\boldsymbol{\gamma}^{(j)}} \quad &
J(\boldsymbol{\gamma}^{(j)}) =
\int_{t_j}^{t_{j+1}}
\left\| x_{\tx{M}}(t) - x(t;\,\boldsymbol{\gamma}^{(j)}) \right\|_G^2
\, \tx{d}t \label{eq:param_id_cost}\\
\text{s.t.} \quad &
\dot{x}(t) = f\!\left(x(t),\, u(t),\, d(t),\, v,\,
\boldsymbol{\gamma}^{(j)},\, \boldsymbol{\theta}\right)
\label{eq:param_id_dyn}\\
& x(t_j) = x_{\tx{M}}(t_j) \label{eq:param_id_init}
\end{align}
\end{subequations}
Each identified parameter vector is associated with the representative
operating condition
\begin{equation}
\boldsymbol{\gamma}^{(j)} \approx
\boldsymbol{\gamma}\!\left(T_{\tx{amb}}^{(j)},\, \bar{x}^{(j)},\,
\bar{I}_{\tx{b}}^{(j)}\right)
\label{eq:gamma_map}
\end{equation}

% -----------------------------------------------------------------------
\section{Model Validation}
\label{sec:ModelValidation}

\begin{figure}[t]
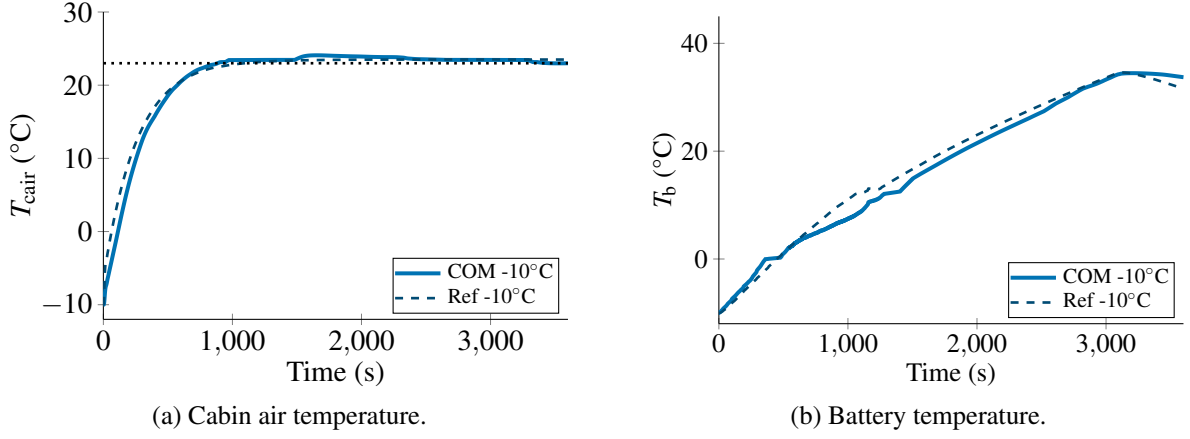

\centering
\begin{subfigure}[t]{0.48\linewidth}
    \centering
    \input{Figures/tikz_validation/val_cabin_air}
    \caption{Cabin air temperature.}
    \label{fig:val_cabin}
\end{subfigure}
\hfill
\begin{subfigure}[t]{0.48\linewidth}
    \centering
    \input{Figures/tikz_validation/val_battery}
    \caption{Battery temperature.}
    \label{fig:val_battery}
\end{subfigure}
\caption{COM (solid) and high-fidelity reference (dashed) trajectories at
  \SI{-10}{\celsius}. Temperatures are shown in degrees Celsius.}
\label{fig:validation}
\end{figure}

The COM is validated against the MathWorks reference over an extended drive
cycle spanning \SI{3600}{\second}. Three cold-climate ambient temperatures
are considered: \SI{-5}{\celsius}, \SI{-7}{\celsius}, and
\SI{-10}{\celsius}.

\begin{table}[ht]
\centering
\caption{Validation RMSE and MAE of the COM against the high-fidelity
  reference. All errors in \si{\celsius}.}
\label{tab:validation_metrics}
\begin{tabular}{lcccccc}
\hline
& \multicolumn{2}{c}{\SI{-10}{\celsius}}
& \multicolumn{2}{c}{\SI{-7}{\celsius}}
& \multicolumn{2}{c}{\SI{-5}{\celsius}} \\
State & RMSE & MAE & RMSE & MAE & RMSE & MAE \\
\hline
$T_{\tx{mot}}$  & 1.60 & 1.35 & 1.70 & 1.46 & 1.43 & 1.18 \\
$T_{\tx{inv}}$  & 1.50 & 1.26 & 1.53 & 1.33 & 1.29 & 1.05 \\
$T_{\tx{dcdc}}$ & 2.12 & 1.61 & 2.00 & 1.62 & 1.83 & 1.30 \\
$T_{\tx{b}}$    & 1.61 & 1.36 & 1.93 & 1.77 & 1.53 & 1.33 \\
$T_{\tx{cair}}$ & 1.04 & 0.58 & 1.45 & 1.14 & 1.01 & 0.81 \\
\hline
\end{tabular}
\end{table}

All five thermal states remain within $2.2\,^{\circ}$C MAE. The cabin air
temperature achieves a MAE below $1.15\,^{\circ}$C across all scenarios.
The battery temperature tracks the reference within $1.8\,^{\circ}$C MAE.

% -----------------------------------------------------------------------
\section{Hybrid Control Architecture}
\label{sec:nmpc}

The proposed control architecture consists of two layers operating at the
same sampling interval $\Delta t$. The upper layer is a rule-based
supervisory controller that determines the discrete system configuration at
each sampling instant. Given this fixed configuration, the lower layer is a
continuous NMPC that optimizes the remaining actuator commands to minimize
thermal energy consumption over a finite prediction horizon.

\subsection{Prediction Model and Hybrid Decomposition}
For use in the NMPC, the continuous-time dynamics are discretized using
fourth-order Runge--Kutta (RK4) integration:
\begin{equation}
x_{k+1}=\Phi\!\left(x_k,\,u_k,\,d_k,\,v_k,\,\boldsymbol{\gamma}_k,\,
\boldsymbol{\theta}_k\right)
\label{eq:discrete_dynamics_full}
\end{equation}
The discrete inputs are determined by the supervisory layer before each
NMPC update and held fixed over the full prediction horizon,
$v_k = v, \; k = 0,\dots,N-1.$
The stage parameter vector at each step is
\begin{equation}
z_k=\begin{bmatrix}
d_k^\top & v^\top & \boldsymbol{\gamma}^\top & \boldsymbol{\theta}^\top
\end{bmatrix}^\top
\label{eq:stage_params}
\end{equation}
so the prediction model simplifies to
\begin{equation}
x_{k+1}=f(x_k,\,u_k;\,z_k)
\label{eq:discrete_dynamics}
\end{equation}

\begin{figure}[!t]
\centering
\includegraphics[width=\linewidth]{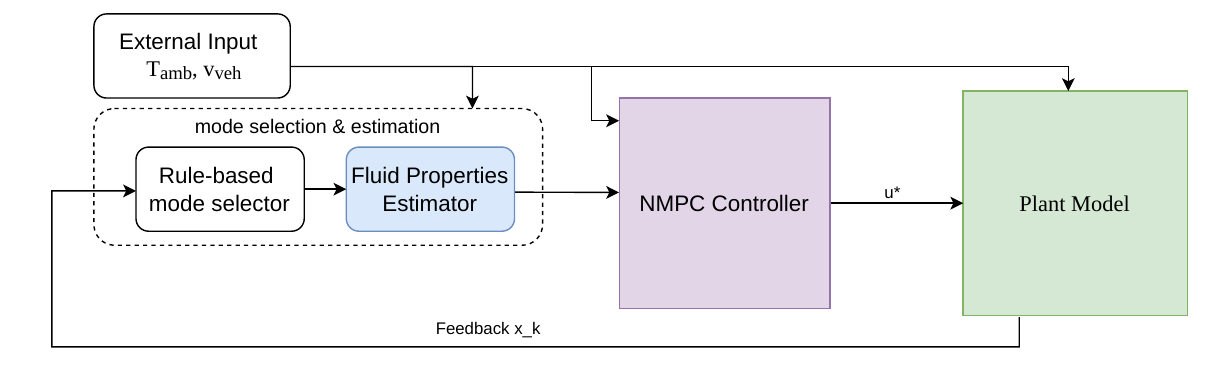}
\caption{Closed-loop control architecture. The rule-based mode selector
  determines discrete supervisory decisions $v$ from external inputs. The
  estimator supplies the current state estimate $\hat{x}_k$, and the NMPC
  applies the optimal input $u_k^*$ to the plant at each sampling instant.}
\label{fig:ControllerArchitecture}
\end{figure}

\subsection{Optimal Control Problem}
At each sampling instant $t$, the controller solves
\begin{subequations}\label{eq:nmpc_problem}
\begin{align}
\min_{\mathbf{x},\mathbf{u},\mathbf{s}} \quad
& \sum_{k=0}^{N-1}\ell(x_k,u_k,z_k) + V_f(x_N)
+ V_{N-1}^{\tx{tie}}(u_{N-1}) \nonumber\\
& + \sum_{k=1}^{N}\!\left(\|W_{x,\ell}s_k^{x,\ell}\|_2^2
+ \|W_{x,u}s_k^{x,u}\|_2^2\right) \nonumber\\
& + \sum_{k=0}^{N-1}\|W_y s_k^y\|_2^2
+ \sum_{k=0}^{N-1}\|W_{\Delta u}s_k^{\Delta u}\|_2^2
\label{eq:nmpc_cost}\\
\text{s.t.}\quad
& x_0 = \hat{x}_t \label{eq:nmpc_ic}\\
& x_{k+1} = f(x_k,u_k;z_k),\quad k=0,\dots,N-1 \label{eq:nmpc_dyn}\\
& x_{\min} \preceq x_k \preceq x_{\max},\quad k=1,\dots,N \label{eq:nmpc_hard_state}\\
& u_{\min} \preceq u_k \preceq u_{\max},\quad k=0,\dots,N-1 \label{eq:nmpc_hard_input}\\
& x_{\tx{pref,lo}} \preceq x_k + s_k^{x,\ell},\quad k=1,\dots,N \label{eq:nmpc_pref_lo}\\
& x_k - s_k^{x,u} \preceq x_{\tx{pref,hi}},\quad k=1,\dots,N \label{eq:nmpc_pref_hi}\\
& g(x_k,u_k;z_k) + s_k^y \succeq 0,\quad k=0,\dots,N-1 \label{eq:nmpc_alg}\\
& {-}\Delta u_{\max} - s_k^{\Delta u} \preceq u_k - u_{k-1}
  \preceq \Delta u_{\max} + s_k^{\Delta u} \label{eq:nmpc_rate}\\
& s_k^{x,\ell} \succeq 0,\quad s_k^{x,u} \succeq 0,\quad k=1,\dots,N \label{eq:nmpc_slacks_x}\\
& s_k^{y} \succeq 0,\quad s_k^{\Delta u} \succeq 0,\quad k=0,\dots,N-1 \label{eq:nmpc_slacks_u}
\end{align}
\end{subequations}

\subsection{Cost Function}
\label{sec:cost-Func}

\subsubsection{Stage Cost}
\begin{align}
\ell(x_k,u_k;z_k)
&= (x_k-x_k^{\tx{ref}})^\top W_{\tx{st}} (x_k-x_k^{\tx{ref}})
+ w_{\tx{pwr}}\,P_{\tx{TEM}}(x_k,u_k;z_k) \nonumber\\
&\quad + u_k^\top R u_k
+ (u_k-u_{k-1})^\top R_{\Delta u,k}(u_k-u_{k-1})
\label{eq:stage_cost}
\end{align}

\subsubsection{Terminal Cost}
\label{sec:terminal-cost}
The quasi-steady terminal pair $(x_{\tx{st}},u_{\tx{st}})$ is obtained by
solving
\begin{equation}
\begin{aligned}
\min_{x,\,u,\,s}\quad
& \lambda_T\!\left(T_{\tx{cair}}-T_{\tx{cair}}^{\tx{ref}}\right)^2
+ \rho_2 \|s\|_2^2 + \rho_u \|u\|_2^2 \\
\text{s.t.}\quad
& x-\Phi(x,u;y_N)=s,\quad
  x_{\min}\preceq x \preceq x_{\max},\quad
  u_{\min}\preceq u \preceq u_{\max}
\end{aligned}
\label{eq:terminal-target}
\end{equation}
The quadratic terminal penalty is
\begin{equation}
V_f(x_N)=(x_N-x_{\tx{st}})^\top P (x_N-x_{\tx{st}})
\label{eq:Vlqr}
\end{equation}
The matrix $P \succ 0$ is obtained from the DARE,
\begin{equation}
P = \mathbf{Q}_P + A^\top P A -
A^\top P B(B^\top P B + \mathbf{R}_P)^{-1} B^\top P A
\label{eq:dare}
\end{equation}
The tie-breaking term on the final predicted input is
\begin{equation}
V_{N-1}^{\tx{tie}}(u_{N-1}) = \beta\,(u_{N-1}-u_{\tx{st}})^\top
R_{\tx{tie}}\,(u_{N-1}-u_{\tx{st}})
\label{eq:Vtie}
\end{equation}

\subsection{Receding Horizon Implementation}

\begin{algorithm}[h]
\caption{Receding-horizon NMPC implementation}
\label{alg:nmpc}
\begin{algorithmic}[1]
\STATE \textbf{Initialization:} Set $t \gets 0$, obtain $\hat{x}_0$,
  initialize $u_{-1}$
\REPEAT
  \STATE Measure or estimate current state $\hat{x}_t$
  \STATE Obtain disturbance preview $\{d_t, \dots, d_{t+N-1}\}$
  \STATE Evaluate $\gamma(T_\tx{amb}, x, I_\tx{b})$ and $\theta$
  \STATE Assemble stage-wise parameter sequence $\{z_k\}_{k=0}^{N-1}$
  \STATE Solve~\eqref{eq:terminal-target} for $(x_{\tx{st}}, u_{\tx{st}})$
  \STATE Linearize at $(x_{\tx{st}}, u_{\tx{st}})$; compute $P$ via
    Riccati fallback
  \STATE Construct warm start by shifting previous optimal solution
  \STATE Solve OCP~\eqref{eq:nmpc_problem} via warm-started interior-point
  \IF{NLP converges}
    \STATE Apply $u_t \gets u_0^\star$
  \ELSE
    \STATE Compute $K = (B^\top P B + R_P)^{-1} B^\top P A$
    \STATE Apply saturated LQR fallback:
      $u_t \leftarrow \mathrm{sat}(u_{\tx{st}} + K(x_{\tx{st}} -
      \hat{x}_t),\; u_{\min},\; u_{\max})$
  \ENDIF
  \STATE Set $t \gets t + \Delta t$
\UNTIL{drive cycle completed}
\end{algorithmic}
\end{algorithm}

% -----------------------------------------------------------------------
\section{Simulation Results and Discussion}
\label{sec:Results}

\subsection{Simulation Setup}
All simulations use the MathWorks Simscape \emph{Electric Vehicle Thermal
Management with Heat Pump} model as the high-fidelity plant. The NMPC is
implemented in MATLAB using CasADi~\cite{Andersson2019} with IPOPT and the
MA97 linear solver. Prediction horizon $N = 30$ steps, sampling time
$\Delta t = 1$\,s. Three cold-climate scenarios are evaluated at
\SI{-5}{\celsius}, \SI{-7}{\celsius}, and \SI{-10}{\celsius}, all starting
from rest at ambient temperature with cabin setpoint
$T_{\tx{cair}}^{\tx{ref}} = \SI{21}{\celsius}$.

\subsection{Thermal State Trajectories}

\begin{figure}[t]
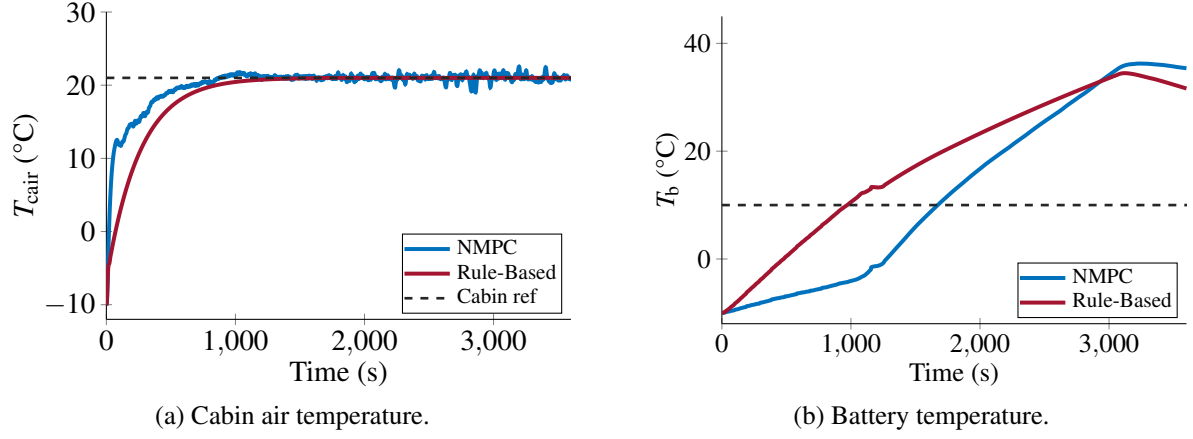

\centering
\begin{subfigure}[t]{0.48\linewidth}
    \centering
    \input{Figures/Controller/nmpc_T10_cabin_temperature}
    \caption{Cabin air temperature.}
    \label{fig:nmpc_T10_cabin}
\end{subfigure}
\hfill
\begin{subfigure}[t]{0.48\linewidth}
    \centering
    \input{Figures/Controller/nmpc_T10_battery_temperature}
    \caption{Battery temperature.}
    \label{fig:nmpc_T10_battery}
\end{subfigure}
\caption{Thermal state trajectories at \SI{-10}{\celsius}: (a) cabin air
  temperature and (b) battery temperature. The NMPC reaches the comfort
  setpoint faster while the battery temperature rises more gradually.}
\label{fig:nmpc_T10_temp}
\end{figure}

\subsection{Energy Consumption}

\begin{table}[h]
\centering
\caption{Total TEM energy consumption over the extended drive cycle.}
\label{tab:energy}
\begin{tabular}{lccc}
\hline
& \SI{-5}{\celsius} & \SI{-7}{\celsius} & \SI{-10}{\celsius} \\
\hline
Baseline (Wh) & 2280 & 2570 & 3020 \\
NMPC (Wh)     & 1820 & 1900 & 2180 \\
Reduction (\%) & 20.2 & 26.1 & 27.8 \\
\hline
\end{tabular}
\end{table}

The NMPC achieves consistent energy reductions ranging from
\SI{20.2}{\percent} at \SI{-5}{\celsius} to \SI{27.8}{\percent} at
\SI{-10}{\celsius}. Two actuators account for the majority of the savings
at \SI{-10}{\celsius}: the auxiliary heater (the NMPC modulates heater power
continuously instead of on/off at maximum) and the compressor (reduced
speed compensated by increased blower flow at lower electrical cost).

\subsection{Computational Performance}
Mean IPOPT solver times were \SI{498.4}{\milli\second},
\SI{577.5}{\milli\second}, and \SI{471.4}{\milli\second} for the three
scenarios. For real-time deployment, dedicated tools such as \texttt{acados}
would further reduce computation time.

% -----------------------------------------------------------------------
\section{Conclusion}
\label{sec:conclusion}

Predictive control of a fully integrated BEV thermal management system
delivers 20--28\% energy savings over rule-based strategies under
cold-climate conditions without sacrificing cabin comfort or violating
component thermal constraints. The adaptive terminal cost---recomputed at
every sampling instant---is the key enabler, particularly near mode
transitions. The publicly available benchmark and open-source
implementation at
\href{https://github.com/PrashantLokur/ThermalEnergyManagementWithHybridControlFramework}{GitHub}
support direct reproducibility.

Future work includes extension to hot-climate cooling-mode operation,
distributed MPC decomposition to reduce per-step computation, and
adaptation of the hybrid control structure to fuel cell vehicles and
second-life battery storage systems.

% -----------------------------------------------------------------------
\section*{Acknowledgment}
This research was partly funded by the Swedish Energy Agency through the
FFI program for Vehicle Research and Innovation (Authority's Dnr
2024-202023, Project No.\ P2024-01000), and by Geely Technology Europe.
During the preparation of this manuscript, the authors used an AI-based
assistance tool for grammar check. The authors carefully reviewed and
revised all suggested edits and take full responsibility for the final
content.

% -----------------------------------------------------------------------
\bibliographystyle{IEEEtran}
\bibliography{references_complete}

@misc{MathWorksEVHeatPumpExample,
  author       = {{MathWorks}},
  title        = {Electric Vehicle Thermal Management with Heat Pump},
  howpublished = {Simscape Fluids example model (Open Model)},
  year         = {2026},
  note         = {Accessed: 2026-02-11}
}

@misc{StrictEnvironmentalRegulations,
  author = {{European Union}},
  title = {Regulation (EU) 2019/631 setting CO$_2$ emission performance standards for new passenger cars and for new light commercial vehicles},
  year = {2019},
  note = {Official Journal of the European Union}
}

@article{RangeAnxietyStudies1,
  author = {Neaimeh, M. and Salisbury, S. D. and Hill, G. A. and Blythe, P. T. and Scoffield, D. R. and Francfort, J. E.},
  title = {Analyzing the usage and evidencing the importance of fast chargers for the adoption of battery electric vehicles},
  journal = {Energy Policy},
  volume = {108},
  pages = {474--486},
  year = {2017},
  doi = {10.1016/j.enpol.2017.06.033}
}

@article{ThermalLoadImpact1,
  author = {Paffumi, E. and De Gennaro, M. and Martini, G. and Manfredi, U.},
  title = {Experimental Test Campaign on a Battery Electric Vehicle: On-Road Test Results (Part 2)},
  journal = {SAE Int. J. Alt. Power},
  volume = {4},
  number = {2},
  pages = {277--292},
  year = {2015},
  doi = {10.4271/2015-01-1166}
}

@article{ThermalLoadImpact2,
  author  = {Lohse-Busch, Henning and Duoba, Michael and Rask, Eric
             and Stutenberg, Kevin and Gowri, Vivek and Slezak, Lee
             and Anderson, David},
  title   = {Ambient Temperature ({20\textdegree F}, {72\textdegree F}
             and {95\textdegree F}) Impact on Fuel and Energy Consumption
             for Several Conventional Vehicles, Hybrid and Plug-In Hybrid
             Electric Vehicles and Battery Electric Vehicle},
  journal = {SAE Technical Paper Series},
  number  = {2013-01-1462},
  year    = {2013},
  doi     = {10.4271/2013-01-1462}
}

@article{MPCforAutomotive1,
  author = {Sciarretta, A. and Guzzella, L.},
  title = {Control of hybrid electric vehicles},
  journal = {IEEE Control Syst. Mag.},
  volume = {27},
  number = {2},
  pages = {60--70},
  year = {2007},
  doi = {10.1109/MCS.2007.338279}
}

@article{MPCforAutomotive2,
  author = {Afram, A. and Janabi-Sharifi, F.},
  title = {Theory and applications of HVAC control systems – A review of model predictive control (MPC)},
  journal = {Building and Environment},
  volume = {72},
  pages = {343--355},
  year = {2014},
  doi = {10.1016/j.buildenv.2013.11.016}
}

@article{HeatPump,
  author = {Meyer, J. J. and Lustbader, J. and Agathocleous, N. and Vespa, A. and Rugh, J. and Titov, G.},
  title = {Range Extension Opportunities While Heating a Battery Electric Vehicle},
  journal = {SAE Technical Paper Series},
  number = {2018-01-0066},
  year = {2018},
  doi = {10.4271/2018-01-0066}
}

@article{Pevec,
  author = {Pevec, D. and Babic, J. and Carvalho, A. and Ghiassi-Farrokhfal, Y. and Ketter, W. and Podobnik, V.},
  title = {A survey-based assessment of how existing and potential electric vehicle owners perceive range anxiety},
  journal = {Journal of Cleaner Production},
  volume = {276},
  pages = {122779},
  year = {2020},
  doi = {10.1016/j.jclepro.2020.122779}
}

@article{9108617,
  author    = {S. Park and C. Ahn},
  title     = {Computationally Efficient Stochastic Model Predictive Controller for Battery Thermal Management of Electric Vehicle},
  journal   = {IEEE Transactions on Vehicular Technology},
  volume    = {69},
  number    = {8},
  pages     = {8407--8419},
  year      = {2020},
  month     = aug,
  doi       = {10.1109/TVT.2020.2999939}
}

@article{zhang2015energy,
  author  = {Zhang, Quansheng and Stockar, Stephanie and Canova, Marcello},
  title   = {Energy-Optimal Control of an Automotive Air Conditioning
             System for Ancillary Load Reduction},
  journal = {IEEE Transactions on Control Systems Technology},
  volume  = {24},
  number  = {1},
  pages   = {67--80},
  year    = {2016},
  doi     = {10.1109/TCST.2015.2418322}
}

@article{lokur2024control,
  title={Control-oriented Model for Thermal Energy Management of Battery Electric Vehicles},
  author={Lokur, Prashant and Murgovski, Nikolce and Larsson, Mikael},
  journal={IEEE Transactions on Vehicular Technology},
  year={2024},
  publisher={IEEE}
}

@article{schaut2019thermal,
  title={Thermal management for the cabin of a battery electric vehicle considering passengers’ comfort},
  author={Schaut, Stefan and Sawodny, Oliver},
  journal={IEEE Transactions on Control Systems Technology},
  volume={28},
  number={4},
  pages={1476--1492},
  year={2019},
  publisher={IEEE}
}

@ARTICLE{Dennis,
  author={Kibalama, Dennis and Liu, Yuxing and Stockar, Stephanie and Canova, Marcello},
  journal={IEEE Control Systems Letters}, 
  title={Model Predictive Control for Automotive Climate Control Systems via Value Function Approximation}, 
  year={2022},
  volume={6},
  number={},
  pages={1820-1825},
  doi={10.1109/LCSYS.2021.3134199}}

@article{XIE2021116084,
title = {An improved intelligent model predictive controller for cooling system of electric vehicle},
journal = {Applied Thermal Engineering},
volume = {182},
pages = {116084},
year = {2021},
issn = {1359-4311},
doi = {https://doi.org/10.1016/j.applthermaleng.2020.116084},
url = {https://doi.org/10.1016/j.applthermaleng.2020.116084},
author = {Yi Xie and Zhaoming Liu and Kuining Li and Jiangyan Liu and Yangjun Zhang and Dan Dan and Cunxue Wu and Pingzhong Wang and Xiaobo Wang}
}

@INPROCEEDINGS{9589323,
  author={Alizadeh, Maryam and Dhale, Sumedh and Emadi, Ali},
  booktitle={IECON 2021 – 47th Annual Conference of the IEEE Industrial Electronics Society}, 
  title={Model Predictive Control of {HVAC} System in a Battery Electric Vehicle with Fan Power Adaptation for Improved Efficiency and Online Estimation of Ambient Temperature}, 
  year={2021},
  volume={},
  number={},
  pages={1-6},
  doi={10.1109/IECON48115.2021.9589323}}

@Article{Fei,
AUTHOR = {Ju, Fei and Murgovski, Nikolce and Zhuang, Weichao and Wang, Liangmo},
TITLE = {Integrated Propulsion and Cabin-Cooling Management for Electric Vehicles},
JOURNAL = {Actuators},
VOLUME = {11},
YEAR = {2022},
NUMBER = {12},
ARTICLE-NUMBER = {356},
URL = {https://www.mdpi.com/2076-0825/11/12/356},
ISSN = {2076-0825},
DOI = {10.3390/act11120356}
}

@article{jeffers2016climate,
  title={Climate control load reduction strategies for electric drive vehicles in cold weather},
  author={Jeffers, Matthew A and Chaney, Larry and Rugh, John P},
  journal={SAE International Journal of Passenger Cars-Mechanical Systems},
  volume={9},
  number={2016-01-0262},
  pages={75--82},
  year={2016}
}

@ARTICLE{7529090,
  author={Lopez-Sanz, Jorge and Ocampo-Martinez, Carlos and Alvarez-Florez, Jesus and Moreno-Eguilaz, Manuel and Ruiz-Mansilla, Rafael and Kalmus, Julian and Gräeber, Manuel and Lux, Gerhard},
  journal={IEEE Transactions on Vehicular Technology}, 
  title={Nonlinear Model Predictive Control for Thermal Management in Plug-in Hybrid Electric Vehicles}, 
  year={2017},
  volume={66},
  number={5},
  pages={3632-3644},
  doi={10.1109/TVT.2016.2597242}}

@article{BAUER2014808,
title = {Thermal and energy battery management optimization in electric vehicles using Pontryagin's maximum principle},
journal = {Journal of Power Sources},
volume = {246},
pages = {808-818},
year = {2014},
issn = {0378-7753},
doi = {https://doi.org/10.1016/j.jpowsour.2013.08.020},
url = {https://www.sciencedirect.com/science/article/pii/S0378775313013591},
author = {Sebastian Bauer and Andre Suchaneck and Fernando {Puente León}}
}

@article{WU2024122090,
title = {Optimal battery thermal management for electric vehicles with battery degradation minimization},
journal = {Applied Energy},
volume = {353},
pages = {122090},
year = {2024},
issn = {0306-2619},
doi = {https://doi.org/10.1016/j.apenergy.2023.122090},
url = {https://www.sciencedirect.com/science/article/pii/S030626192301454X},
author = {Yue Wu and Zhiwu Huang and Dongjun Li and Heng Li and Jun Peng and Daniel Stroe and Ziyou Song}
}

@misc{cengel2003heat,
  title={Heat Transfer A Practical Approach},
  author={Cengel, Yunus A},
  year={2003},
  publisher={McGraw-Hill}
}

@article{hajidavalloo2023nmpc,
  title={NMPC-based integrated thermal management of battery and cabin for electric vehicles in cold weather conditions},
  author={Hajidavalloo, Mohammad R and Chen, Jun and Hu, Qiuhao and Song, Ziyou and Yin, Xunyuan and Li, Zhaojian},
  journal={IEEE Transactions on Intelligent Vehicles},
  volume={8},
  number={9},
  pages={4208--4222},
  year={2023},
  publisher={IEEE}
}

@article{CHEN202244,
title = {Control-oriented Model of HVAC and Battery Cooling Systems in Electric Vehicles},
journal = {IFAC-PapersOnLine},
volume = {55},
number = {37},
pages = {44-49},
year = {2022},
note = {2nd Modeling, Estimation and Control Conference MECC 2022},
issn = {2405-8963},
doi = {https://doi.org/10.1016/j.ifacol.2022.11.159},
url = {https://www.sciencedirect.com/science/article/pii/S2405896322028026},
author = {Youyi Chen and Kyoung Hyun Kwak and Jaewoong Kim and Dewey D. Jung and Youngki Kim}
}

@article{bemporad1999control,
  title={Control of systems integrating logic, dynamics, and constraints},
  author={Bemporad, Alberto and Morari, Manfred},
  journal={Automatica},
  volume={35},
  number={3},
  pages={407--427},
  year={1999},
  publisher={Elsevier}
}

@article{Bell2014,
  author  = {Bell, Ian H. and Wronski, Jorrit and Quoilin, Sylvain 
             and Lemort, Vincent},
  title   = {Pure and Pseudo-pure Fluid Thermophysical Property 
             Evaluation and the Open-Source Thermophysical Property 
             Library {CoolProp}},
  journal = {Industrial \& Engineering Chemistry Research},
  volume  = {53},
  number  = {6},
  pages   = {2498--2508},
  year    = {2014},
  doi     = {10.1021/ie4033999}
}

@Article{Andersson2019,
  author = {Joel A E Andersson and Joris Gillis and Greg Horn
            and James B Rawlings and Moritz Diehl},
  title = {{CasADi} -- {A} software framework for nonlinear optimization
           and optimal control},
  journal = {Mathematical Programming Computation},
  volume = {11},
  number = {1},
  pages = {1--36},
  year = {2019},
  publisher = {Springer},
  doi = {10.1007/s12532-018-0139-4}
}

@article{janardhanan2025energy,
  title={Energy-efficient wheel torque distribution for heavy electric vehicles with adaptive model predictive control and control allocation},
  author={Janardhanan, Sachin and Persson, Jonas and Jonasson, Mats and Jacobson, Bengt and Gelso, Esteban R and Henderson, Leon},
  journal={IEEE Open Journal of Vehicular Technology},
  year={2025},
  publisher={IEEE}
}

% -----------------------------------------------------------------------
\section*{Author Biographies}

\textbf{Prashant Lokur} is an Industrial PhD student jointly affiliated
with the Department of Advanced Motion Systems and Energy at Geely
Technology Europe and the Department of Electrical Engineering, Division of
Systems and Control, Chalmers University of Technology, Gothenburg, Sweden.
He received the B.Eng.\ degree in Mechanical Engineering from Visvesvaraya
Technological University, India, in 2012, and the M.Sc.\ degree in Systems,
Control, and Mechatronics from Chalmers University of Technology in 2015.
His current research focuses on optimal control of thermal energy management
systems for battery electric vehicles.

\textbf{Nikolce Murgovski} is a Professor with the Department of
Electrical Engineering, Division of Systems and Control, Chalmers
University of Technology. He received the M.S.\ degree in Software
Engineering from University West, Sweden, in 2007, the M.S.\ degree in
Applied Physics, and the PhD degree in Systems and Control from Chalmers in
2007 and 2012, respectively. His research interests are in optimization,
optimal control, and online learning within electromobility and autonomous
driving.

\end{document}